%% file: paper.tex
\documentclass{article}

\usepackage{microtype}
\usepackage{graphicx}
\usepackage{subcaption}
\usepackage{booktabs}
\usepackage{hyperref}

\usepackage[preprint]{icml2026}

\usepackage{amsmath}
\usepackage{amssymb}
\usepackage{mathtools}
\usepackage{amsthm}
\usepackage{algorithm}
\usepackage{algorithmic}
\usepackage{listings}
\usepackage{xcolor}
\usepackage{tikz}
\usepackage{fontawesome5}
\usetikzlibrary{positioning, arrows.meta, shapes.geometric, fit, calc, backgrounds}
\usepackage[capitalize,noabbrev]{cleveref}
\usepackage[textsize=tiny]{todonotes}
\usepackage{placeins}
\usepackage{afterpage}

\theoremstyle{plain}

\theoremstyle{definition}

\theoremstyle{remark}

\definecolor{leankw}{RGB}{0,80,160}
\definecolor{leancmt}{RGB}{120,120,120}
\lstdefinelanguage{Lean4}{
  morekeywords={theorem,lemma,def,structure,inductive,instance,variable,
                fun,let,by,exact,intro,intros,apply,rw,simp,have,show,
                refine,calc,cases,induction,unfold,rfl,if,then,else,
                match,with,where,namespace,end,open,import,axiom,sorry,admit,
                obtain,rcases,classical,haveI},
  sensitive=true,
  morecomment=[l]{--},
  morecomment=[s]{/-}{-/},
  morestring=[b]",
  extendedchars=true,
  literate=
    {α}{{$\alpha$}}1 {β}{{$\beta$}}1 {γ}{{$\gamma$}}1 {δ}{{$\delta$}}1
    {ε}{{$\varepsilon$}}1 {ζ}{{$\zeta$}}1 {η}{{$\eta$}}1 {θ}{{$\theta$}}1
    {λ}{{$\lambda$}}1 {μ}{{$\mu$}}1 {π}{{$\pi$}}1 {ρ}{{$\rho$}}1
    {σ}{{$\sigma$}}1 {τ}{{$\tau$}}1 {φ}{{$\varphi$}}1 {ψ}{{$\psi$}}1
    {ω}{{$\omega$}}1 {Γ}{{$\Gamma$}}1 {Δ}{{$\Delta$}}1 {Σ}{{$\Sigma$}}1
    {Π}{{$\Pi$}}1 {Λ}{{$\Lambda$}}1
    {∀}{{$\forall$}}1 {∃}{{$\exists$}}1 {∈}{{$\in$}}1 {∉}{{$\notin$}}1
    {∧}{{$\wedge$}}1 {∨}{{$\vee$}}1 {¬}{{$\neg$}}1 {⊥}{{$\bot$}}1 {⊤}{{$\top$}}1
    {→}{{$\to$}}1 {←}{{$\leftarrow$}}1 {↦}{{$\mapsto$}}1 {⇒}{{$\Rightarrow$}}1
    {≃}{{$\simeq$}}1 {≅}{{$\cong$}}1 {≈}{{$\approx$}}1
    {≤}{{$\leq$}}1 {≥}{{$\geq$}}1 {≠}{{$\neq$}}1 {≡}{{$\equiv$}}1
    {⟨}{{$\langle$}}1 {⟩}{{$\rangle$}}1
    {·}{{$\cdot$}}1 {∘}{{$\circ$}}1 {×}{{$\times$}}1
    {ℕ}{{$\mathbb{N}$}}1 {ℝ}{{$\mathbb{R}$}}1 {ℤ}{{$\mathbb{Z}$}}1
    {ℚ}{{$\mathbb{Q}$}}1 {ℂ}{{$\mathbb{C}$}}1 {ℓ}{{$\ell$}}1
    {ä}{{\"a}}1 {ö}{{\"o}}1 {ü}{{\"u}}1 {Ä}{{\"A}}1 {Ö}{{\"O}}1 {Ü}{{\"U}}1
    {ß}{{\ss}}1 {é}{{\'e}}1 {è}{{\`e}}1 {á}{{\'a}}1 {à}{{\`a}}1
    {í}{{\'i}}1 {ó}{{\'o}}1 {ú}{{\'u}}1 {ñ}{{\~n}}1
    {⁻}{{$^{-}$}}1 {¹}{{$^{1}$}}1 {²}{{$^{2}$}}1 {³}{{$^{3}$}}1 {ⁿ}{{$^{n}$}}1
    {–}{{--}}1 {—}{{---}}1 {…}{{\ldots}}1 {≃}{{$\simeq$}}1
}
\icmltitlerunning{MerLean-Prover}

\begin{document}

\twocolumn[
  \icmltitle{MerLean-Prover: A Recursive Looping Harness for Lean~4 Theorem Proving}

  \icmlsetsymbol{equal}{*}

  \begin{icmlauthorlist}
    \icmlauthor{Jinzheng Li}{equal,JL}
    \icmlauthor{Zeru Zhu}{equal,ZZ}
    \icmlauthor{Yuanjie Ren}{equal,YR}
  \end{icmlauthorlist}
  \icmlaffiliation{JL}{Northeastern University}
  \icmlaffiliation{ZZ}{Stony Brook University}
  \icmlaffiliation{YR}{Massachusetts Institute of Technology}
  \icmlcorrespondingauthor{Jinzheng Li}{li.jinzh@northeastern.edu}
  \icmlcorrespondingauthor{Zeru Zhu}{zeru.zhu@stonybrook.edu}
  \icmlcorrespondingauthor{Yuanjie Ren}{yuanjie@mit.edu}

  \icmlkeywords{theorem proving, autoformalization, Lean 4, agentic systems, AI for math}

  \vskip 0.3in
]

\printAffiliationsAndNotice{}

\begin{abstract}
\textsc{MerLean-Prover} is an end-to-end Lean~4 theorem prover that replaces \texttt{sorry} declarations with kernel-checkable proofs. It is built from three agent types (Planning, Check, and Lean) composed by a recursive outer loop whose unit of revision is the proof plan itself, and uses no fine-tuning, no custom RL objective, and no theorem-specific scaffolding. On \textsc{FormalQualBench}, a benchmark of 23 PhD-qualifying-exam theorems, \textsc{MerLean-Prover} solves \textbf{10/23}, surpassing the strongest published open-source baseline (OpenGauss, 8/23). On Putnam~2025, the same harness closes \textbf{12/12} with substantially lower total wall-clock than the next-best system that closes the full set. The harness also transfers to smaller models: Sonnet closes all four tested \textsc{FormalQualBench} problems, and Haiku closes the two short ones. These results suggest that harness design is a central factor in end-to-end Lean~4 theorem proving, alongside raw model capability, and that a relatively simple harness can already be effective.
\end{abstract}

\afterpage{%
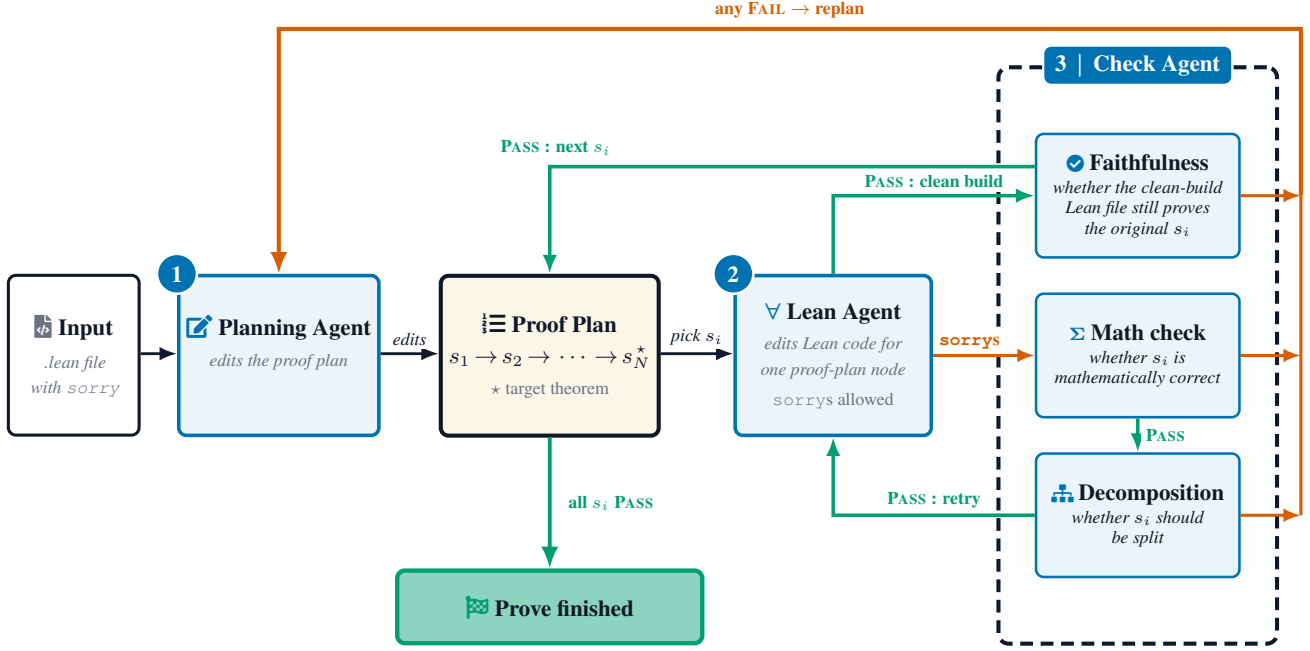
\begin{figure*}[!t]
\centering
\definecolor{ink}{HTML}{111827}
\definecolor{muted}{HTML}{6B7280}
\definecolor{accent}{HTML}{B8501E}
\definecolor{artifill}{HTML}{FAF6E8}
\definecolor{nodeformalized}{HTML}{0072B2} 
\definecolor{nodered}{HTML}{D55E00}        
\colorlet{agentblue}{nodeformalized}
\colorlet{agentfill}{nodeformalized!8}
\definecolor{passgreen}{HTML}{009E73}     
\colorlet{failred}{nodered}
\resizebox{\textwidth}{!}{%
\begin{tikzpicture}[
  every node/.append style={align=center, text=ink, font=\footnotesize},
  box/.style    ={rectangle, draw=ink, line width=1.2pt, rounded corners=3pt,
                  minimum height=22mm, inner sep=3pt},
  inp/.style    ={box, minimum width=18mm, fill=white},
  upd/.style    ={box, minimum width=27mm, fill=agentfill, draw=agentblue,
                  line width=1.4pt},
  stmt/.style   ={box, minimum width=30mm, fill=artifill, line width=1.6pt,
                  rounded corners=3pt},
  wcf/.style    ={box, minimum width=27mm, fill=agentfill, draw=agentblue,
                  line width=1.4pt},
  chksub/.style ={rectangle, draw=agentblue, line width=1.0pt, rounded corners=3pt,
                  minimum width=27mm, minimum height=16mm, fill=agentfill,
                  inner sep=1.5pt, text=ink, font=\scriptsize},
  vrb/.style    ={box, minimum width=42mm, minimum height=10mm,
                  fill=artifill, line width=1.4pt, rounded corners=3pt},
  stepbadge/.style={circle, fill=agentblue, text=white, font=\footnotesize\bfseries,
                    inner sep=0pt, minimum size=5.5mm, draw=white, line width=1pt},
  fwdar/.style   ={-{Latex[length=2.4mm,width=1.9mm]}, line width=1.1pt, draw=ink},
  innerar/.style ={-{Latex[length=2.4mm,width=1.9mm]}, line width=1.1pt, draw=ink},
  passar/.style  ={-{Latex[length=2.6mm,width=2.0mm]}, line width=1.4pt, draw=passgreen},
  recar/.style   ={-{Latex[length=3mm,width=2.4mm]},   line width=1.6pt, draw=failred},
  failar/.style  ={-{Latex[length=2.4mm,width=1.9mm]}, line width=1.2pt, draw=failred},
  exitar/.style  ={-{Latex[length=3mm,width=2.4mm]},   line width=1.5pt, draw=passgreen},
  lbl/.style     ={font=\scriptsize\itshape, text=ink,
                   fill=white, inner sep=1.5pt},
  passlbl/.style ={font=\scriptsize\bfseries, text=passgreen,
                   fill=white, inner sep=1.5pt},
  reclbl/.style  ={font=\scriptsize\bfseries, text=failred,
                   fill=white, inner sep=1.5pt},
]

\node[inp, anchor=north west] (input) at (0,0)
  {{\color{muted}\faFileCode}\;{\small\bfseries Input}\\[3pt]
   {\color{muted}\itshape\scriptsize\,.lean file}\\
   {\color{muted}\itshape\scriptsize\,with \texttt{sorry}}};

\node[upd, anchor=north west] (update) at ([xshift=5mm]input.north east)
  {{\color{agentblue}\faEdit}\;{\small\bfseries Planning Agent}\\[2pt]
   {\color{muted}\itshape\scriptsize edits the proof plan}\\[2pt]};

\node[stmt, anchor=north west] (planjson) at ([xshift=8mm]update.north east)
  {{\color{ink}\faListOl}\;{\small\bfseries Proof Plan}\\[2pt]
   $s_1 \,{\to}\, s_2 \,{\to}\, \cdots \,{\to}\, s_N^{\,\star}$\\[2pt]
   {\color{muted}\scriptsize $\star$ target theorem}};

\node[wcf, anchor=north west] (compile) at ([xshift=10mm]planjson.north east)
  {{\color{agentblue}$\boldsymbol{\forall}$}\;{\small\bfseries Lean Agent}\\[2pt]
   {\color{muted}\itshape\scriptsize edits Lean code for}\\
   {\color{muted}\itshape\scriptsize one proof-plan node}\\[2pt]
   {\color{muted}\scriptsize\texttt{sorry}s allowed}};

\node[chksub, anchor=west, minimum height=17mm, minimum width=28mm]
  (faith) at ([xshift=14mm, yshift=22mm]compile.east)
  {{\color{agentblue}\faCheckCircle}\;{\small\bfseries Faithfulness}\\[1pt]
   {\itshape whether the clean-build}\\
   {\itshape Lean file still proves}\\
   {\itshape the original $s_i$}};
\node[chksub, anchor=west, minimum height=17mm, minimum width=28mm]
  (math) at ([yshift=-22mm]faith.west)
  {{\color{agentblue}$\boldsymbol{\Sigma}$}\;{\small\bfseries Math check}\\[1pt]
   {\itshape whether $s_i$ is}\\
   {\itshape mathematically correct}};
\node[chksub, anchor=west, minimum height=17mm, minimum width=28mm]
  (decomp) at ([yshift=-22mm]math.west)
  {{\color{agentblue}\faSitemap}\;{\small\bfseries Decomposition}\\[1pt]
   {\itshape whether $s_i$ should}\\
   {\itshape be split}};

\begin{scope}[on background layer]
  \node[draw=ink, line width=1.6pt, dash pattern=on 5pt off 3pt,
        inner xsep=5mm, inner ysep=9mm, rounded corners=4pt,
        fill=agentfill!15,
        fit=(faith)(math)(decomp)] (chkbox) {};
\end{scope}

\node[fill=agentblue, text=white, font=\footnotesize\bfseries,
      inner sep=3pt, rounded corners=3pt, draw=white, line width=0.5pt]
  at (chkbox.north)
  {\,3\, \textbar\, Check Agent\,};

\node[stepbadge] at (update.north west)  {1};
\node[stepbadge] at (compile.north west) {2};

\node[vrb, anchor=north, fill=passgreen!45, draw=passgreen, line width=1.7pt]
  at ([yshift=-18mm]planjson.south) (verdict)
  {{\color{passgreen}\faFlagCheckered}\;{\small\bfseries\color{ink} Prove finished}};

\coordinate (mainY) at (0, -11mm);

\draw[fwdar] (input.east  |- mainY) -- (update.west  |- mainY);
\draw[fwdar] (update.east |- mainY) -- node[lbl, midway, above=0.5mm]{edits}
             (planjson.west |- mainY);
\draw[fwdar] (planjson.east |- mainY) -- node[lbl, midway, above=0.5mm]{pick $s_i$}
             (compile.west |- mainY);

\draw[passar] (compile.north) |- (faith.west);
\node[passlbl, anchor=south, yshift=0.5mm]
  at ($(compile.north |- faith.west)!0.5!(faith.west)$) {\textsc{Pass} : clean build};

\draw[failar] (compile.east) -- (math.west);
\node[anchor=south, yshift=0.5mm, xshift=-2mm,
      font=\scriptsize\bfseries, text=failred, fill=white, inner sep=1.5pt]
  at ($(compile.east)!0.5!(math.west)$) {\texttt{sorry}s};

\coordinate (faith_pass)  at ([yshift=4mm]faith.west);
\draw[passar] (faith_pass) -| (planjson.north);
\node[passlbl, anchor=south east] at ([xshift=9mm, yshift=1mm]planjson.north |- faith_pass)
   {\textsc{Pass} : next $s_i$};

\draw[passar] (math.south) -- (decomp.north);
\node[passlbl, anchor=west, xshift=0.5mm] at ($(math.south)!0.5!(decomp.north)$)
   {\textsc{Pass}};

\coordinate (decomp_pass_out) at (decomp.west);
\coordinate (decomp_pass_l)   at (compile.south |- decomp_pass_out);
\draw[passar] (decomp_pass_out) -- (decomp_pass_l) -- (compile.south);
\node[passlbl, anchor=south, yshift=0.5mm] at ($(decomp_pass_out)!0.5!(decomp_pass_l)$)
   {\textsc{Pass} : retry};

\coordinate (faith_fail)  at (faith.east);
\coordinate (math_fail)   at (math.east);
\coordinate (decomp_fail) at (decomp.east);

\coordinate (fail_bus)      at ([xshift=3mm]chkbox.east);
\coordinate (faith_fail_c)  at (faith_fail  -| fail_bus);
\coordinate (math_fail_c)   at (math_fail   -| fail_bus);
\coordinate (decomp_fail_c) at (decomp_fail -| fail_bus);

\draw[failar] (faith_fail)  -- (faith_fail_c);
\draw[failar] (math_fail)   -- (math_fail_c);
\draw[failar] (decomp_fail) -- (decomp_fail_c);
\draw[line width=1.2pt, draw=failred] (faith_fail_c) -- (decomp_fail_c);

\coordinate (chk_above) at ([yshift=5mm]chkbox.north);
\coordinate (fail_top)  at (fail_bus     |- chk_above);
\coordinate (upd_top)   at (update.north |- chk_above);
\draw[recar] (faith_fail_c) -- (fail_top) -- (upd_top) -- (update.north);
\node[reclbl, anchor=south,yshift = 1mm] at ($(fail_top)!0.5!(upd_top)$)
   {any \textsc{Fail} $\to$ replan};

\draw[exitar] (planjson.south) -- (verdict.north);
\node[passlbl, anchor=west, xshift=2mm]
   at ($(planjson.south)!0.5!(verdict.north)$)
   {all $s_i$ \textsc{Pass}};

\end{tikzpicture}%
}
\caption{\textsc{MerLean-Prover}'s recursive looping architecture. The shared proof plan is the unit of revision. The Lean Agent works on one statement at a time; fresh Check Agent instances each answer one decision question. \emph{Math}: whether $s_i$ is mathematically correct; \emph{Decomposition}: whether $s_i$ should be split; \emph{Faithfulness}: whether the clean-build Lean file still proves the original $s_i$. A clean build goes to Faithfulness before the loop advances to the next $s_i$; if the Lean file has any \texttt{sorry} in it, the Lean file is checked by Math, then Decomposition, before the Lean Agent resumes. Any \textsc{Fail} invokes the Planning Agent for a replan, invalidating affected downstream statements and restarting from the top of the revised plan. The success exit is available only when all plan statements pass and the target theorem still matches the original Lean signature.}
\label{fig:arch}
\end{figure*}%
}
\section{Introduction}
Language-model agents can now compile Lean~4, parse error messages, and iterate~\cite{polu2020generativelanguagemodelingautomated,han2022pact,yang2023leandojo}. A growing family of agentic harnesses (including OpenGauss~\cite{mathinc2025opengauss}, Numina-Lean-Agent~\cite{liu2026numinaleanagentopengeneralagentic}, Seed-Prover~1.5~\cite{chen2025seedprover15masteringundergraduatelevel}, Hilbert~\cite{varambally2025hilbertrecursivelybuildingformal}, Ax-Prover~\cite{breen2025axproverdeepreasoningagentic}, Archon~\cite{ju2026automatedconjectureresolutionformal}, and unmodified general-purpose coding agents such as Claude Code, Codex, and opencode) has achieved strong performance across formal mathematics benchmarks. Earlier targets are now largely closed: Seed-Prover~1.5~\cite{chen2025seedprover15masteringundergraduatelevel} reportedly saturates MiniF2F~\cite{zheng2021minif2f}, PutnamBench~\cite{tsoukalas2024putnam} is nearing closure, and Numina-Lean-Agent~\cite{liu2026numinaleanagentopengeneralagentic} and Axiom~\cite{axiom2025} have fully formalized Putnam~2025. PhD-qualifying-exam difficulty~\cite{mathinc2025formalqualbench} remains an open challenge; at this scale, the central difficulty shifts from individual proof attempts to long-horizon agent behavior.

Long proof attempts expose two well-documented LLM weaknesses at once. \emph{Context pressure}: long-context recall degrades non-monotonically with input length~\cite{liu2024lostmiddle}. \emph{Constraint pressure}: instruction-following accuracy decays as simultaneous constraints accumulate~\cite{jaroslawicz2025instructionsllmsfollowonce}. A graduate-level proof attempt accumulates dozens of dependencies and a long error history, while every invocation must also preserve the original signature, avoid \texttt{axiom}, stay tactical, and not weaken hypotheses. In theorem proving, errors have high downstream cost: a single slip can change the target statement, admit an inadmissible axiom, or invalidate downstream proofs.

These weaknesses suggest a harness-level response: keep global state outside the prompt, give each agent invocation one focused task, and let the harness enforce global constraints. The proof plan externalizes memory; each invocation sees a local proof object and one local objective; recursion revisits the plan after failures without packing the entire proof history into one context. \textsc{MerLean-Prover} instantiates this principle as an end-to-end harness that replaces \texttt{sorry} declarations in a Lean~4 file with kernel-checkable proofs of the same signatures, using three agent types composed by a recursive outer loop (\cref{sec:arch}). We demonstrate this loop end-to-end on an example (\cref{sec:examples}). The empirical claim of this paper is that this relatively simple architecture can solve a substantial subset of graduate-level formalization problems without fine-tuning or theorem-specific scaffolding.

In this work, \textsc{MerLean-Prover} achieves \textbf{10/23} on \textsc{FormalQualBench}, with nine solves within the matched 4-hour budget and one in an extended 4h40m run, surpassing the published open-source state of the art (OpenGauss, 8/23). On the Putnam~2025 slice, the same harness closes \textbf{12/12} and is the fastest system on 8 of 12 problems. We additionally report a repeated-run stability study showing that closures are reproducible across attempts, and a Sonnet/Haiku variant study showing that the harness still produces kernel-audited proofs when run with smaller models (\cref{sec:results}).

\section{Architecture}
\label{sec:arch}

\Cref{fig:arch} sketches the full pipeline. The system accepts a \texttt{.lean} file with one or more \texttt{sorry} declarations, generates a proof plan, and runs a recursive compile-and-revise loop until the target theorem is closed or an external stop condition is reached. We describe this shared object first, then the three single-objective agent roles, and finally the recursive loop. The key design choice is to keep global state in the proof plan while giving each agent invocation one local task.

\paragraph{Shared proof plan.} The proof plan is the only global object shared across invocations. Conceptually, it is a topologically sorted list of proof statements: helper nodes appear before the statements that depend on them, and each node contains an informal statement, a proof sketch, its dependencies, and its completion status. For the target theorem, the plan stores an \emph{anchor}: the original Lean declaration with its \texttt{sorry} body. This anchor records the exact theorem signature---name, parameters, assumptions, and conclusion---that the final proof must preserve. When a node changes, the harness invalidates the downstream statements that depend on it, so later invocations see a consistent local view of the revised plan.

\paragraph{Three agent types.} We use \emph{agent type} to mean a role class with a fixed authority boundary: the Planning Agent may edit the proof plan, the Lean Agent may edit Lean code, and the Check Agent is read-only and answers one true/false question. Within each type, different task prompts define different variants, but the authority boundary remains the same. Every invocation is given exactly one objective with the minimum context it needs. This three-role decomposition follows the harness-agent paradigm, where the harness coordinates planning, code execution, and verification across separate LLM calls~\cite{ning2026codeasagentharness}.
\begin{itemize}\itemsep1pt
  \item The \textbf{Planning Agent} proposes or edits the proof plan. It is invoked in four situations: initial plan construction from the input file, and three recovery cases triggered by the check agents: faithfulness failure, mathematical-correction failure, and decomposition-driven replanning. Its output is either the initial plan or a proposed plan diff.
  \item The \textbf{Lean Agent} writes Lean for one statement in the proof plan and then iteratively fixes it until \texttt{lake build} passes or exhausts its compile budget. \texttt{sorry} placeholders are allowed in its output---intermediate scaffolding is acceptable during this step, since any surviving \texttt{sorry} is caught downstream by the harness and handled by the Math check or the Decomposition check. It reports either a build result or a compile-exhaustion signal.
  \item The \textbf{Check Agent} audits outputs produced by the other agents. Each invocation runs independently from the producer session, sees the output rather than the producer's chain-of-thought, and returns one true/false verdict. Its variants ask three decision questions: whether a statement is mathematically correct as stated, whether a stuck statement should be split into smaller sub-statements, and whether a clean-build Lean file still proves the original statement.
\end{itemize}
No single prompt is asked to carry all global constraints at once; instead, the harness enforces signature preservation, axiom-freedom, and faithfulness across the loop.

\paragraph{Recursive loop.} The outer loop repeatedly walks the proof plan in dependency order and applies the Lean Agent to each open statement. The Lean Agent's build result and the relevant Check Agent verdicts determine the next step: close the statement, try the same statement again, or ask the Planning Agent to revise the plan. The normal stopping point is closure of the current plan, including the anchored target; if the run reaches an external time limit before this point, the proof is not counted as solved. A clean build is sent to the Faithfulness check; if it passes, the loop advances to the next statement. When the Lean Agent cannot close the current statement, the Math check first asks whether the statement is mathematically correct as stated. If the statement is sound, the Decomposition check asks whether it should be split into smaller sub-statements. If the answer is no, the Lean Agent retries the same statement; if the answer is yes, the Planning Agent turns the failure into a plan diff. Failed Math and Faithfulness checks also ask the Planning Agent to revise the plan. When the diff is accepted, the harness invalidates affected downstream statements and restarts from the top of the updated plan.

A run is counted as solved only when every statement in the current plan is closed, the target theorem still has the original Lean signature, and the final proof contains no forbidden \texttt{sorry}, \texttt{admit}, or user-introduced \texttt{axiom}. For benchmark reporting, an external goal audit checks this condition; if it fails, we count the run as unfinished.

\paragraph{Design scope.} We did not fine-tune any model, introduce a custom RL objective, engineer theorem-specific prompts, or include manually written proof sketches. The same harness, prompt set, and default model (\texttt{claude-opus-4-7}) are used on every problem. Initial planning, Lean writing and repair, and the Math, Decomposition, and Faithfulness checks are all delegated to standard calls to a general-purpose LLM with single-objective prompts. This input format is an interface choice rather than a conceptual restriction: broader inputs, such as a natural-language proof in \LaTeX{}, can be handled by a preprocessing converter that produces the same Lean-with-\texttt{sorry} interface~\cite{ren2026merleanautoformalization}. Such converters are separate from the recursive proof loop and are outside the present evaluation.

\FloatBarrier

\section{Example Discussion}
\label{sec:examples}

Before turning to aggregate numbers, one worked example makes the recursive loop concrete. We use \textsc{BurnsidePrimeDegreeTheorem} because its trace exercises all three recovery cases from \cref{sec:arch}: faithfulness failure, mathematical-correction failure, and decomposition-driven replanning. The anchored target is the main theorem, and node labels below match \cref{fig:burnside-evolution}, with the full node map in \cref{app:burnside-node-map}.

\input{_fig2_tikz.tex}

\paragraph{BurnsidePrimeDegreeTheorem.} The Planning Agent's initial plan contains 6 nodes: the anchored target (node~\#1) and five prerequisites (nodes~\#2--\#6). The Lean Agent then processes the prerequisites, closes four of them, and leaves node~\#2. Node~\#2 reaches a clean \texttt{lake build}, but the Faithfulness check rejects it as weakened at frame~6, a faithfulness failure. The Planning Agent revises the plan in response: it rewrites node~\#2's statement to remove the weakening (a mathematical-correction step) and introduces node~\#7 as a new helper to carry the missing content (decomposition-driven replanning). The Lean Agent then works on node~\#7 until it exhausts its repair budget at frame~16, at which point the Decomposition check votes for a split, and the Planning Agent replaces node~\#7 with smaller helpers, another instance of decomposition-driven replanning. The same pattern recurs at nodes~\#8, \#13, and \#23. Mathematical-correction failure arises explicitly during the node~\#13 replan: the Math check identifies that one of the helpers under node~\#13 has a signature missing a required hypothesis (without it the lemma would be mathematically false), so the Planning Agent rewrites the statement before introducing further sub-helpers. Each recovery case becomes a plan revision; the plan grows from 6 to 12 nodes, then to 21, then to 32. By frame~51 the node set has stabilized. The Lean Agent then closes the remaining nodes bottom-up, and frame~63 closes the anchored target~\#1. The final goal check reports that the anchored declaration is present and that no \texttt{sorry}, \texttt{admit}, or \texttt{axiom} remains in the proof closure. The closed Lean source for nodes~\#1 and \#2 is reproduced in \cref{app:lean-sample}.

\paragraph{Proof cone structure.} We use \emph{proof cone} as a visual term for the dependency shape produced by this recursion: many small helper nodes feed fewer intermediate nodes, and those feed the anchored target. In this run, the cone is not an asymptotic claim; it is the concrete record of how a 6-node plan becomes a 32-node plan as decomposition-driven replanning is invoked four times on hard nodes. This is the key behavior of the harness. It does not ask one invocation to solve the whole hard step; it asks the Planning Agent to make the next proof statement smaller, then lets the Lean Agent close the smaller pieces. The faithfulness and mathematical-correction cases play complementary roles: faithfulness failure catches the rare clean-build Lean file that has drifted from the anchored signature before that drift propagates downstream, and mathematical-correction failure catches helper statements whose signature is internally inconsistent before the Lean Agent burns its budget on a target that cannot be true.

\section{Results and Discussion}
\label{sec:results}

\paragraph{Setup.} We evaluate on \textsc{FormalQualBench} as the primary benchmark. The remaining experiments report a \textsc{PutnamBench} slice, repeated-run stability, and Sonnet/Haiku model variants. A problem counts as solved only if the produced Lean builds cleanly, preserves the original theorem signature, and passes the transitive \texttt{\#print axioms} audit with \texttt{sorryAx} excluded and only $\{\texttt{propext},\,\texttt{Quot.sound},\,\texttt{Classical.choice}\}$ permitted. Unless stated otherwise, the budget is 4 hours of wall-clock time per problem. Cost is reported in USD of LLM API spend. \textsc{MerLean-Prover} uses \texttt{claude-opus-4-7} with default settings; Lean~4 v4.28.0 with Mathlib~\cite{mathlib2020} v4.28.0.

\begin{table}[t]
\caption{\textsc{FormalQualBench} closure rate. Baseline numbers are from \cite{mathinc2025formalqualbench}; *Aristotle results are unvalidated. $^{\dagger}$Nine of the ten \textsc{MerLean-Prover} solves close within the 4-hour budget; one solve closes in an extended 4h40m run.}
\label{tab:main}
\centering
\footnotesize
\setlength{\tabcolsep}{4pt}
\begin{tabular}{lc}
\toprule
System & Solved / 23 \\
\midrule
\textbf{\textsc{MerLean-Prover} (ours)} & \textbf{10}$^{\dagger}$ \\
OpenGauss               & 8 \\
Aristotle*              & 6 \\
Claude Code (Skills)    & 5 \\
Codex                   & 5 \\
opencode (Opus)         & 5 \\
Claude Code             & 4 \\
Claude Code (MCP)       & 3 \\
Codex (Skills + MCP)    & 3 \\
\bottomrule
\end{tabular}
\end{table}

\begin{table}[t]
\caption{Cost and wall-clock time for the ten \textsc{MerLean-Prover} solves on \textsc{FormalQualBench}. Problem IDs are mapped to theorem names in \cref{app:fqb-id-map}. Token costs are computed from model-provider usage logs. n/r marks per-problem costs the leaderboard does not report. --- means OpenGauss did not solve the theorem. $^{\dagger}$ID~9 is the only solve outside the 4-hour budget.}
\label{tab:fqb-cost-time}
\centering
\footnotesize
\setlength{\tabcolsep}{5pt}
\begin{tabular}{@{}r|cccc@{}}
\toprule
& \multicolumn{2}{c}{\textsc{MerLean-Prover} (ours)} & \multicolumn{2}{c}{OpenGauss} \\
\cmidrule(lr){2-3}\cmidrule(lr){4-5}
ID & Time & Cost & Time & Cost \\
\midrule
1   & 3h41m                      & \$220.50          & \textbf{2h13m} & n/r              \\
2   & \textbf{3h12m}             & \textbf{\$184.27} & ---            & ---              \\
3   & 1h52m                      & \$127.28          & \textbf{1h50m} & \textbf{\$27.68} \\
4   & \textbf{11m}               & \$10.05           & 12m            & \textbf{\$2.54}  \\
5   & \textbf{1h24m}             & \$78.21           & 1h52m          & n/r              \\
6   & \textbf{2h13m}             & \$118.33          & 3h20m          & n/r              \\
7   & \textbf{16m}               & \$13.53           & 24m            & \textbf{\$9.99}  \\
8   & \textbf{1h01m}             & \$62.16           & 1h34m          & \textbf{\$15.70} \\
9   & \textbf{4h40m}$^{\dagger}$ & \textbf{\$273.66} & ---            & ---              \\
10  & \textbf{1h23m}             & \$95.47           & 2h50m          & n/r              \\
\bottomrule
\end{tabular}
\end{table}

\paragraph{\textsc{FormalQualBench}.} \Cref{tab:main} reports the primary leaderboard comparison, while \cref{tab:fqb-cost-time} gives the cost and time profile of the solved problems. The problem IDs in \cref{tab:fqb-cost-time} are mapped to full theorem names in \cref{app:fqb-id-map}; IDs 11--23 are the remaining benchmark problems not closed by \textsc{MerLean-Prover}. \textsc{MerLean-Prover} solves \textbf{10/23}, surpassing OpenGauss (the strongest published open-source baseline) and every other published system on the leaderboard. Nine solves close within the published 4-hour wall-clock budget; ID~9 is the one extended run and closes in 4h40m. Across the ten solves, total LLM API spend is \$1{,}183.45 and total wall-clock time is 19h54m, for averages of \$118.35 and 1h59m per solved problem. The cost-to-time ratio is higher than OpenGauss's because the recursive loop accumulates a large volume of prompt-cache reads and writes across replans: cache traffic is much faster to process than fresh generation, but every token is still billed, so the harness stays within a tight wall-clock budget while the per-problem dollar cost grows with plan size. Every solve passes the \textsc{Comparator}-style transitive \texttt{\#print axioms} audit, i.e.\ is axiom-free in the kernel sense. This performance is achieved without fine-tuning or theorem-specific scaffolding.

\begin{table}[t]
\caption{Per-problem wall-clock time (minutes) on Putnam~2025. Comparison columns for Aristotle~\cite{achim2025aristotleimolevelautomatedtheorem}, Seed-Prover~1.5~\cite{chen2025seedprover15masteringundergraduatelevel}, Axiom~\cite{axiom2025}, and Numina-Lean-Agent (NLA)~\cite{liu2026numinaleanagentopengeneralagentic} are reproduced from Table~3 of~\cite{liu2026numinaleanagentopengeneralagentic}; ``--'' marks problems not solved.}
\label{tab:putnam}
\centering
\footnotesize
\setlength{\tabcolsep}{4pt}
\begin{tabular}{@{}l|rrrrr@{}}
\toprule
Prob. & Arist. & Seed~1.5 & Axiom & NLA & \textbf{Ours} \\
\midrule
A1 & 30           & 60           & 110 & 97           & \textbf{27}  \\
A2 & 60           & \textbf{30}  & 180 & \textbf{30}  & 31           \\
A3 & \textbf{30}  & 120          & 165 & 44           & 44           \\
A4 & 180          & 240          & 107 & 169          & \textbf{38}  \\
A5 & --           & --           & 518 & 2040         & \textbf{235} \\
A6 & 60           & 240          & 259 & 89           & \textbf{25}  \\
B1 & 150          & 540          & 270 & \textbf{55}  & 161          \\
B2 & \textbf{25}  & 360          & 65  & 142          & 55           \\
B3 & 40           & 30           & 43  & 30           & \textbf{16}  \\
B4 & --           & 120          & 112 & 308          & \textbf{53}  \\
B5 & 420          & 240          & 254 & 88           & \textbf{43}  \\
B6 & 180          & 180          & 494 & 797          & \textbf{61}  \\
\midrule
Solved & 10/12 & 11/12 & 12/12 & 12/12 & \textbf{12/12} \\
Total  & ---   & ---   & 2577  & 3889  & \textbf{789}   \\
\bottomrule
\end{tabular}
\end{table}

\paragraph{\textsc{PutnamBench}.} On the Putnam~2025 slice, \textsc{MerLean-Prover} closes \textbf{12/12} (\cref{tab:putnam}), matching the per-set count of Numina-Lean-Agent~\cite{liu2026numinaleanagentopengeneralagentic} and Axiom while exceeding Aristotle~\cite{achim2025aristotleimolevelautomatedtheorem} (10/12) and Seed-Prover~1.5~\cite{chen2025seedprover15masteringundergraduatelevel} (11/12). It is also the fastest system on 8 of the 12 problems, and the lowest total wall-clock among systems that close the full set: 789 minutes against Axiom's 2{,}577 and Numina-Lean-Agent's 3{,}889. The slowest \textsc{MerLean-Prover} solve (A5) closes in 235 minutes against Numina's 2040 and Axiom's 518. Every solve passes the same kernel-level \texttt{\#print axioms} audit.

\begin{table}[t]
\caption{Repeated-run timing stability on four selected \textsc{FormalQualBench} problems. Each problem is run eight times and all attempts produce clean proofs. Time, statements, and minutes per statement report mean~$\pm$~std.; median and min--max report wall-clock time. Problem IDs follow \cref{app:fqb-id-map}.}
\label{tab:stability}
\centering
\footnotesize
\setlength{\tabcolsep}{2pt}
\begin{tabular}{@{}r|cccccc@{}}
\toprule
ID & Time & Median & Min--max & Stmts & Min/stmt \\
\midrule
1 & 3.83$\pm$0.39h & 3.85h & 3.27--4.40h & 16.1$\pm$3.4 & 14.9$\pm$3.5 \\
3 & 1.49$\pm$0.63h & 1.22h & 0.86--2.77h & 7.3$\pm$2.8  & 12.5$\pm$3.0 \\
4 & 0.22$\pm$0.04h & 0.21h & 0.17--0.29h & 2.0$\pm$0.0  & 6.5$\pm$1.1 \\
7 & 0.31$\pm$0.06h & 0.30h & 0.25--0.44h & 2.9$\pm$1.0  & 6.8$\pm$1.2 \\
\bottomrule
\end{tabular}
\end{table}

\paragraph{Stability.} Because the harness is stochastic and recursively revises its own proof plan, a single successful run does not show whether the behavior is typical. We therefore run \textsc{MerLean-Prover} eight times on each of four selected problems from \textsc{FormalQualBench}, independent from \cref{tab:fqb-cost-time}, and report the timing distribution in \cref{tab:stability}. All runs produce clean proofs. The variation is mainly explained by two quantities: the number of statements in the final proof plan and the average time needed to close each statement. In the reported samples, both quantities vary only moderately across runs, giving a reasonably stable timing profile.

In a separate pilot run on ID~1 (Banach--Stone), outside the eight repeated runs reported in \cref{tab:stability}, we observed one genuine outlier: the final plan grew to 58 statements and took 17 hours, while still closing the target theorem. We hypothesize that repeated replanning can occasionally select a proof plan that is poorly supported by existing Mathlib lemmas, causing the loop to introduce too many helper statements before it converges. A comprehensive explanation of this behavior remains under investigation.

\begin{table}[t]
\caption{Per-problem wall-clock time and cost across model variants (\texttt{claude-opus-4-7}, \texttt{claude-sonnet-4-6}, \texttt{claude-haiku-4-5}) on the same four \textsc{FormalQualBench} problems as \cref{tab:stability}. Problem IDs follow \cref{app:fqb-id-map}. --- marks runs that did not finish; Haiku did not close IDs~1 or 3 within a 12-hour wall-clock budget.}
\label{tab:model-variants}
\centering
\footnotesize
\setlength{\tabcolsep}{4pt}
\begin{tabular}{@{}r|cccccc@{}}
\toprule
   & \multicolumn{2}{c}{Opus} & \multicolumn{2}{c}{Sonnet} & \multicolumn{2}{c}{Haiku} \\
\cmidrule(lr){2-3}\cmidrule(lr){4-5}\cmidrule(lr){6-7}
ID & Time & Cost & Time & Cost & Time & Cost \\
\midrule
1 & 3h41m & \$220.50 & 11h00m & \$111.89 & ---   & ---     \\
3 & 1h52m & \$127.28 & 5h09m  & \$32.33  & ---   & ---     \\
4 & 11m   & \$10.05  & 17m    & \$3.37   & 29m   & \$2.25  \\
7 & 16m   & \$13.53  & 1h00m  & \$12.00  & 4h34m & \$26.24 \\
\bottomrule
\end{tabular}
\end{table}

\paragraph{Sonnet and Haiku.} \Cref{tab:model-variants} compares the default Opus run with Sonnet and Haiku variants on the four \textsc{FormalQualBench} problems also used for stability analysis. Sonnet closes all four under the same harness, with cost 2--4$\times$ lower than Opus on three of the four problems (and roughly on par on ID~7) at a 1.5--4$\times$ wall-clock penalty. Haiku closes only the two short problems (IDs~4 and 7) and does not finish IDs~1 or 3 within a 12-hour budget. On ID~7, the cost pattern reverses: Haiku spends 4h34m for \$26.24 against Opus's 16m for \$13.53. Haiku produces a clean proof but only after decomposing the target into a nine-statement plan against Opus's three, and the accumulated token traffic over that longer plan outweighs Haiku's 15$\times$ lower per-token rate. We read this experiment as evidence that the harness itself, rather than the strongest model alone, contributes substantially to the decomposition; on small problems the cheaper models reduce cost, but on harder ones a weaker model can lengthen the proof plan enough to offset the per-token cost advantage.

\section{Conclusion}
\label{sec:conclusion}

\textsc{MerLean-Prover} is built around one architectural commitment: \emph{one objective per agent invocation}. With three agent types and a recursive outer loop, it matches the strongest published open-source result on \textsc{FormalQualBench}, without fine-tuning, a custom RL objective, or theorem-specific scaffolding. These results suggest that harness design is a central factor in end-to-end Lean~4 theorem proving, alongside raw model capability, and that a relatively simple harness can already be effective.

\paragraph{Strength.} \emph{Model leverage.} The main strength of \textsc{MerLean-Prover} is that it turns a general-purpose reasoning model into a competitive Lean prover through harness design alone. Without fine-tuning, a custom RL objective, or theorem-specific scaffolding, it closes 10/23 \textsc{FormalQualBench} problems and outperforms the strongest published open-source baseline. \emph{Repeatability.} On the four selected \textsc{FormalQualBench} problems in the stability study, all eight repeated runs close cleanly. This suggests that the recursive loop is not merely finding a single lucky trajectory: when the problem is within reach, the harness can repeatedly build a complete proof plan and close it. \emph{Model flexibility.} The same harness also transfers to smaller models: Sonnet closes all four shared problems, and Haiku closes the two short ones. This suggests that part of the capability comes from the harness structure itself, not only from the default Opus model.

\paragraph{Weakness.} \emph{Model-capability floor.} Because the harness does not use theorem-specific scaffolding or fine-tuning, its capability still depends strongly on the underlying reasoning model. The model must propose useful decompositions, write Lean repairs, and judge mathematical correctness and faithfulness. If the model is too weak on any of these roles, the recursive loop loses much of its advantage. Haiku sits near this boundary: it closes the short problems in \cref{tab:model-variants} but does not finish the harder ones within 12 hours. \emph{Mathlib coverage.} The system relies strongly on support from Mathlib. The initial proof plan, generated helper statements, and final Lean repairs all depend on having relevant library facts available and usable. On hard targets, the loop may require facts that are missing from Mathlib, or facts that exist but are difficult for the model to locate and use. This limitation is common to Lean-based theorem-proving systems that rely on the existing library. \emph{Cost.} The recursive loop is expensive: failed compile attempts, replanning steps, and repeated reads of a growing proof plan all create additional model calls and token traffic. This makes the current system costly to scale and difficult to use in cost-sensitive settings.

\paragraph{Forward look.} Three extensions are natural. (i) The current implementation is not optimized; engineering work to reduce redundant compile attempts, repeated planning, and accumulated cache traffic could lower per-problem cost without changing the design. (ii) Because the harness still closes problems with smaller models (\cref{tab:model-variants}), pairing it with an open-source Lean-specific model such as Leanstral~\cite{mistral2026leanstral} is a natural direction for reducing cost further. (iii) Failed decompositions can be used as signals for missing or under-documented Mathlib components.

\bibliography{paper}
\bibliographystyle{icml2026}

\newpage
\appendix

\onecolumn

\section{\textsc{FormalQualBench} Problem ID Map}
\label{app:fqb-id-map}

\begin{table}[h]
\caption{Problem IDs used in \cref{tab:fqb-cost-time}. IDs 1--10 are the problems closed by \textsc{MerLean-Prover}; IDs 11--23 are the remaining \textsc{FormalQualBench} problems.}
\label{tab:fqb-id-map}
\centering
\footnotesize
\setlength{\tabcolsep}{5pt}
\begin{tabular}{@{}r|lr|l@{}}
\toprule
\multicolumn{2}{c}{Closed by \textsc{MerLean-Prover}} &
\multicolumn{2}{c}{Remaining benchmark problems} \\
\cmidrule(r){1-2}\cmidrule(l){3-4}
ID & Theorem & ID & Theorem \\
\midrule
1  & \texttt{BanachStoneTheorem}                 & 11 & \texttt{BorsukUlamTheorem} \\
2  & \texttt{BurnsidePrimeDegreeTheorem}         & 12 & \texttt{CollatzMapAlmostBoundedValues} \\
3  & \texttt{ColorfulCaratheodoryTheorem}        & 13 & \texttt{ErdosDiscrepancyProblem} \\
4  & \texttt{DeBruijnErdos}                      & 14 & \texttt{GreenTaoTheorem} \\
5  & \texttt{DLOQuantifierElimination}           & 15 & \texttt{Hilbert17thProblem} \\
6  & \texttt{GleasonKahaneZelazkoTheorem}        & 16 & \texttt{JordanCycleTheorem} \\
7  & \texttt{JordanDerangementTheorem}           & 17 & \texttt{KakeyaTheorem3D} \\
8  & \texttt{ParisHarringtonPrinciple}           & 18 & \texttt{MaynardTaoBoundedPrimeGaps} \\
9  & \texttt{RungeTheorem}                       & 19 & \texttt{PontryaginDuality} \\
10 & \texttt{VonNeumannDoubleCommutantTheorem}   & 20 & \texttt{QuillenSuslinTheorem} \\
   &                                               & 21 & \texttt{SchauderFixedPointTheorem} \\
   &                                               & 22 & \texttt{SkolemMahlerLechTheorem} \\
   &                                               & 23 & \texttt{TernaryGoldbachTheorem} \\
\bottomrule
\end{tabular}
\end{table}

\section{BurnsidePrimeDegreeTheorem Node Map}
\label{app:burnside-node-map}

\begin{table}[h]
\caption{Display node labels used in \cref{fig:burnside-evolution}.}
\label{tab:burnside-node-map}
\centering
\scriptsize
\setlength{\tabcolsep}{2pt}
\begin{tabular}{@{}rlr@{\hspace{0.35cm}}rlr@{}}
\toprule
Node & Statement id & Frame & Node & Statement id & Frame \\
\midrule
\#1  & \texttt{Thm\_MainTheorem}                       & 0  & \#17 & \texttt{Lem\_PiShiftImageEqShiftImage}           & 24 \\
\#2  & \texttt{Lem\_BurnsideDichotomyCore}             & 0  & \#18 & \texttt{Lem\_PolynomialEvalAffine}               & 24 \\
\#3  & \texttt{Lem\_CyclicSubgroupActsRegularly}       & 0  & \#19 & \texttt{Lem\_VandermondePowerSumsContradiction}  & 24 \\
\#4  & \texttt{Lem\_ExistsElementOfPrimeOrder}         & 0  & \#20 & \texttt{Lem\_PolyShiftIdentity}                  & 24 \\
\#5  & \texttt{Lem\_PrimeOrderElementIsCycle}          & 0  & \#21 & \texttt{Lem\_PowerSumsBinomialExpansion}         & 24 \\
\#6  & \texttt{Lem\_PrimeDvdGroupOrder}                & 0  & \#22 & \texttt{Lem\_KeyDegreeInequality}                & 33 \\
\#7  & \texttt{Lem\_MuellerAffineConjugation}          & 11 & \#23 & \texttt{Lem\_CoeffAtNwMinusROfLhs}               & 33 \\
\#8  & \texttt{Lem\_AffinePermOfDifferenceSetPreserving} & 16 & \#24 & \texttt{Lem\_SumCompPowEqOfPerm}                 & 33 \\
\#9  & \texttt{Lem\_NotTwoTransitiveYieldsDifferenceSet} & 16 & \#25 & \texttt{Lem\_LagrangeInterpolantLeadingCoeffNeZero} & 33 \\
\#10 & \texttt{Lem\_TauOrbitEquivConjTau}              & 16 & \#26 & \texttt{Lem\_SumCompPowEqOfShiftIdentity}        & 33 \\
\#11 & \texttt{Def\_TauOrbitEquiv}                     & 16 & \#27 & \texttt{Lem\_NatChooseNotDvdBySmallPrime}        & 33 \\
\#12 & \texttt{Lem\_AffineConjugatesTranslation}       & 16 & \#28 & \texttt{Lem\_PolyEqHigherCoeffZero}              & 33 \\
\#13 & \texttt{Lem\_PowerSumsAllZeroForSmallK}         & 23 & \#29 & \texttt{Lem\_RhsLowerTermsNatDegreeBound}        & 33 \\
\#14 & \texttt{Lem\_PowerSumPiTranslate}               & 24 & \#30 & \texttt{Lem\_CoeffSumGCompXAddCuVanishLow}       & 46 \\
\#15 & \texttt{Lem\_DifferenceSetWLOGSmall}            & 24 & \#31 & \texttt{Lem\_LhsEqSumFwCompXAddCu}               & 46 \\
\#16 & \texttt{Lem\_LagrangeInterpolantOfPerm}         & 24 & \#32 & \texttt{Lem\_CoeffSumGCompXAddCuExpansion}       & 46 \\
\bottomrule
\end{tabular}
\end{table}

\section{Sample Closed Lean Source: Nodes \#1 and \#2 of the BurnsidePrimeDegreeTheorem Trace}
\label{app:lean-sample}

The two listings below reproduce the closed Lean source emitted by \textsc{MerLean-Prover} for the first two nodes of the worked example in \cref{sec:examples}: node~\#1 is the anchored target \texttt{Thm\_MainTheorem}, and node~\#2 is the helper \texttt{Lem\_BurnsideDichotomyCore} after the faithfulness-driven rewrite at frame~11. Each file builds under \texttt{lake build} and passes the transitive \texttt{\#print axioms} audit with no \texttt{sorry}, \texttt{admit}, or user-introduced \texttt{axiom}.

\paragraph{Node \#1: \texttt{Thm\_MainTheorem}.} The anchored target, closed at frame~63 once the cone below has stabilized. The proof case-splits on 2-pretransitivity and delegates the non-trivial direction to node~\#2.

\begin{lstlisting}[language=Lean4]
import Mathlib.GroupTheory.GroupAction.MultipleTransitivity
import FormalQualBench.BurnsidePrimeDegreeTheorem.Auxiliary.BurnsideDichotomyCore

namespace BurnsidePrimeDegreeTheorem
open MulAction

/-- Burnside's theorem on transitive permutation groups of prime
degree: a transitive permutation group of prime degree is either
2-transitive, or it has a normal regular subgroup. -/
theorem MainTheorem
    {α : Type*} [Fintype α]
    {G : Subgroup (Equiv.Perm α)}
    (htrans : IsPretransitive G α)
    (hp : (Fintype.card α).Prime) :
    IsMultiplyPretransitive G α 2 ∨
      ∃ N : Subgroup G, N.Normal ∧ IsPretransitive N α ∧
        ∀ a : α, MulAction.stabilizer N a = ⊥ := by
  classical
  rcases Classical.em (IsMultiplyPretransitive G α 2) with h2 | hNot2
  · exact Or.inl h2
  · haveI : IsPretransitive G α := htrans
    exact Or.inr (burnside_dichotomy_core hp hNot2)

end BurnsidePrimeDegreeTheorem
\end{lstlisting}

\paragraph{Node \#2: \texttt{Lem\_BurnsideDichotomyCore}.} The substantive direction of Burnside's prime-degree theorem, in the form Müller~(2005) gives. The Planning Agent's initial plan contained a weaker statement of this node, which the Faithfulness check rejected at frame~11; the listing below is the post-rewrite form that ultimately closes through node~\#8 and its descendants. The three \texttt{FormalQualBench} imports correspond to the three outgoing edges of node~\#2 in \cref{fig:burnside-evolution}. The file is shown in full: the module docstring is the human-readable plan that the Planning Agent committed to before any compile attempts; the two helper theorems \texttt{cyclicSubgroup\_normal} and \texttt{burnside\_dichotomy\_core\_card} are bookkeeping that the Lean Agent introduced inside the node's file (without inflating the plan-level node count) for use by the main lemma and by the cardinality statement of Burnside's theorem.

\begin{lstlisting}[language=Lean4]
/-
Copyright (c) 2026 MerLean Autoresearch. All rights reserved.
Released under MIT license.
-/
import Mathlib.GroupTheory.Perm.Basic
import Mathlib.GroupTheory.GroupAction.Basic
import Mathlib.GroupTheory.GroupAction.Transitive
import Mathlib.GroupTheory.GroupAction.MultipleTransitivity
import Mathlib.GroupTheory.OrderOfElement
import Mathlib.Algebra.Group.Subgroup.Basic
import Mathlib.Algebra.Group.Subgroup.ZPowers.Basic
import Mathlib.Data.Nat.Prime.Basic
import Mathlib.Data.Fintype.Card
import FormalQualBench.BurnsidePrimeDegreeTheorem.Auxiliary.ExistsElementOfPrimeOrder
import FormalQualBench.BurnsidePrimeDegreeTheorem.Auxiliary.CyclicSubgroupActsRegularly
import FormalQualBench.BurnsidePrimeDegreeTheorem.Auxiliary.MuellerAffineConjugation

/-!
# Burnside dichotomy core

Let `α` be a finite type with `p := Fintype.card α` prime, and let
`G ≤ Equiv.Perm α` act pretransitively on `α`. Assume `G` is not 2-pretransitive.
Then there exists a subgroup `N ≤ G` such that:

1. `N` is normal in `G`,
2. `N` acts pretransitively on `α`,
3. for every `x : α`, the stabilizer in `N` of `x` is trivial (`= ⊥`).

These conditions together state that `N` acts regularly on `α`; in particular
`Nat.card N = p`.

This is the substantive direction of Burnside's classical 1911 theorem on
transitive permutation groups of prime degree, in the elementary form proved
by P. Müller, *Permutation groups of prime degree, a quick proof of Burnside's
theorem*, Arch. Math. 85 (2005), 15–17.

## Strategy

* **Step 1.** Extract `τ : G` with `orderOf τ = p` via
  `exists_element_of_prime_order`.
* **Step 2.** Take `N := Subgroup.zpowers τ`, the cyclic subgroup of `G`
  generated by `τ`.
* **Step 3.** Pretransitivity and trivial stabilizers come from
  `cyclicSubgroup_isPretransitive` and `cyclicSubgroup_stabilizer_eq_bot`.
* **Step 4.** Normality follows from `mueller_affine_conjugation`: for every
  `g : G`, `g τ g⁻¹ ∈ Subgroup.zpowers τ`. We extend to all integer powers
  using `conj_zpow` and `Subgroup.zpow_mem`.

## Main result

* `BurnsidePrimeDegreeTheorem.burnside_dichotomy_core` : the existence of `N`
  with all four properties.
-/

namespace BurnsidePrimeDegreeTheorem

open MulAction

variable {α : Type*} [Fintype α] [DecidableEq α]
variable {G : Subgroup (Equiv.Perm α)}

section BurnsideDichotomy

/-- **Normality of the cyclic subgroup generated by `τ`.**

Under the hypotheses of `mueller_affine_conjugation`, the cyclic subgroup
`Subgroup.zpowers τ ≤ G` is normal in `G`.

The argument: every element of `Subgroup.zpowers τ` is `τ ^ k` for some
`k : ℤ`. For any `g : G`, we have `g * τ ^ k * g⁻¹ = (g * τ * g⁻¹) ^ k` by
`conj_zpow`. Since `g * τ * g⁻¹ ∈ Subgroup.zpowers τ` (by Müller's affine
conjugation lemma), and subgroups are closed under integer powers
(`Subgroup.zpow_mem`), the conjugate lies back in `Subgroup.zpowers τ`. -/
theorem cyclicSubgroup_normal
    [MulAction.IsPretransitive G α]
    (hp : (Fintype.card α).Prime)
    (hNot2 : ¬ MulAction.IsMultiplyPretransitive G α 2)
    (τ : G)
    (hord : orderOf τ = Fintype.card α) :
    (Subgroup.zpowers τ).Normal := by
  refine ⟨?_⟩
  intro h hh g
  -- h ∈ Subgroup.zpowers τ, so h = τ ^ k for some k : ℤ.
  rw [Subgroup.mem_zpowers_iff] at hh
  obtain ⟨k, hk⟩ := hh
  -- Substitute h = τ ^ k.
  rw [← hk]
  -- g * τ^k * g⁻¹ = (g * τ * g⁻¹)^k.
  rw [← conj_zpow]
  -- Now (g * τ * g⁻¹) ∈ Subgroup.zpowers τ by Müller's lemma.
  have hconj : g * τ * g⁻¹ ∈ Subgroup.zpowers τ :=
    mueller_affine_conjugation hp hNot2 τ hord g
  -- Subgroups are closed under integer powers.
  exact Subgroup.zpow_mem _ hconj k

/-- **Burnside dichotomy core.**

Let `α` be a finite type with `p := Fintype.card α` prime, and let
`G ≤ Equiv.Perm α` act pretransitively on `α`. Assume `G` is not
2-pretransitive on `α`. Then there exists a subgroup `N ≤ G` such that:

1. `N` is normal in `G`,
2. `N` acts pretransitively on `α`,
3. for every `x : α`, the stabilizer in `N` of `x` is trivial.

These conditions state that `N` acts regularly on `α`. -/
theorem burnside_dichotomy_core
    [MulAction.IsPretransitive G α]
    (hp : (Fintype.card α).Prime)
    (hNot2 : ¬ MulAction.IsMultiplyPretransitive G α 2) :
    ∃ N : Subgroup G,
      N.Normal ∧
      MulAction.IsPretransitive N α ∧
      ∀ x : α, MulAction.stabilizer N x = ⊥ := by
  -- Step 1: extract τ : G with orderOf τ = p.
  obtain ⟨τ, hord⟩ := exists_element_of_prime_order (G := G) hp
  -- Step 2: take N := Subgroup.zpowers τ.
  refine ⟨Subgroup.zpowers τ, ?_, ?_, ?_⟩
  · -- Normality, by `cyclicSubgroup_normal`.
    exact cyclicSubgroup_normal hp hNot2 τ hord
  · -- Pretransitivity, by `cyclicSubgroup_isPretransitive`.
    exact cyclicSubgroup_isPretransitive τ hp hord
  · -- Trivial stabilizers, by `cyclicSubgroup_stabilizer_eq_bot`.
    intro x
    exact cyclicSubgroup_stabilizer_eq_bot τ hp hord x

/-- **Cardinality of the regular subgroup.**

Under the same hypotheses as `burnside_dichotomy_core`, the regular subgroup
`N := Subgroup.zpowers τ` (with `τ` of order `p`) has `Nat.card N = p`. This
follows from `Nat.card_zpowers` and `orderOf τ = p`. -/
theorem burnside_dichotomy_core_card
    [MulAction.IsPretransitive G α]
    (hp : (Fintype.card α).Prime)
    (hNot2 : ¬ MulAction.IsMultiplyPretransitive G α 2) :
    ∃ N : Subgroup G,
      N.Normal ∧
      MulAction.IsPretransitive N α ∧
      (∀ x : α, MulAction.stabilizer N x = ⊥) ∧
      Nat.card N = Fintype.card α := by
  obtain ⟨τ, hord⟩ := exists_element_of_prime_order (G := G) hp
  refine ⟨Subgroup.zpowers τ, ?_, ?_, ?_, ?_⟩
  · exact cyclicSubgroup_normal hp hNot2 τ hord
  · exact cyclicSubgroup_isPretransitive τ hp hord
  · intro x
    exact cyclicSubgroup_stabilizer_eq_bot τ hp hord x
  · rw [Nat.card_zpowers]
    exact hord

end BurnsideDichotomy

end BurnsidePrimeDegreeTheorem
\end{lstlisting}

\end{document}

%% file: _fig2_tikz.tex
\begin{figure*}[!t]
\centering
\definecolor{nodeformalized}{HTML}{3B5B92} 
\definecolor{nodegray}{HTML}{B4B7BD}
\definecolor{nodered}{HTML}{B65F45}        
\definecolor{nodeink}{HTML}{1A1A1F}
\definecolor{nodefaint}{HTML}{8B8E96}
\def\rad{0.95}
\newlength{\burnsideBaseNodeRadius}
\newlength{\burnsideNodeRadius}
\setlength{\burnsideBaseNodeRadius}{1.25mm}
\setlength{\burnsideNodeRadius}{\rad\burnsideBaseNodeRadius}
\pgfmathsetmacro{\burnsideNodeFontSize}{4.6*\rad}
\def\d{1.1}
\def\xscale{1.06}
\def\edgecolshift{1.65}
\def\rowheight{45.0}
\def\panelheight{90.0}
\def\targetrowmargin{4.5}
\def\toprowgraphshift{1.5}
\def\bottomrowgraphshift{10.5}
\begin{tikzpicture}[
  x=\d mm, y=\d mm,
  every node/.append style={inner sep=0pt},
  dagnode/.style={circle, draw=nodeink, line width=0.3pt, text=white,
                  minimum size=\dimexpr2\burnsideNodeRadius\relax,
                  font=\fontsize{\burnsideNodeFontSize}{\burnsideNodeFontSize}\selectfont\bfseries},
  ndg/.style={dagnode, fill=nodeformalized},
  ndp/.style={dagnode, fill=nodegray},
  ndr/.style={dagnode, fill=nodered},
  ed/.style={-, line width=0.18pt, draw=nodefaint},
  panel/.style={draw=nodeink!55, line width=0.35pt},
  frbadge/.style={anchor=north west, font=\scriptsize\bfseries, text=nodeink},
  status/.style={anchor=north east, font=\scriptsize, text=nodeink},
]
\draw[panel] (0.00,0.00) rectangle (154.00,-\panelheight);
\draw[panel] (51.33,0.00) -- (51.33,-\panelheight);
\draw[panel] (102.67,0.00) -- (102.67,-\panelheight);
\draw[panel] (0.00,-\rowheight) -- (154.00,-\rowheight);
\begin{scope}[shift={(\edgecolshift,0)}, shift={(0,\toprowgraphshift)}, shift={(24.00,0)}, xscale=\xscale, shift={(-24.00,0)}]
\draw[ed] (24.01,-6.00) -- (24.01,-18.00);
\draw[ed] (24.01,-18.00) -- (14.01,-30.00);
\draw[ed] (24.01,-18.00) -- (34.01,-30.00);
\draw[ed] (14.01,-30.00) -- (14.01,-42.00);
\draw[ed] (34.01,-30.00) -- (34.01,-42.00);
\node[ndp] at (14.01,-42.00) {5};
\node[ndp] at (34.01,-42.00) {6};
\node[ndp] at (14.01,-30.00) {3};
\node[ndp] at (34.01,-30.00) {4};
\node[ndp] at (24.01,-18.00) {2};
\node[ndp] at (24.01,-6.00) {1};
\end{scope}
\node[frbadge] at (1.20,-1.20) {frame~0};
\node[status]  at (50.13,-1.20) {0/6 formalized};
\begin{scope}[shift={(0,\toprowgraphshift)}, shift={(77.00,0)}, xscale=\xscale, shift={(-77.00,0)}]
\draw[ed] (77.00,-6.00) -- (77.00,-18.00);
\draw[ed] (77.00,-18.00) -- (67.00,-30.00);
\draw[ed] (77.00,-18.00) -- (87.00,-30.00);
\draw[ed] (67.00,-30.00) -- (67.00,-42.00);
\draw[ed] (87.00,-30.00) -- (87.00,-42.00);
\node[ndg] at (67.00,-42.00) {5};
\node[ndg] at (87.00,-42.00) {6};
\node[ndg] at (67.00,-30.00) {3};
\node[ndg] at (87.00,-30.00) {4};
\node[ndr] at (77.00,-18.00) {2};
\node[ndp] at (77.00,-6.00) {1};
\end{scope}
\node[frbadge] at (52.53,-1.20) {frame~6};
\node[status]  at (101.47,-1.20) {4/6 formalized};
\begin{scope}[shift={(-\edgecolshift,0)}, shift={(0,\toprowgraphshift)}, shift={(130.00,0)}, xscale=\xscale, shift={(-130.00,0)}]
\draw[ed] (129.99,-6.00) -- (129.99,-12.00);
\draw[ed] (129.99,-12.00) -- (119.99,-36.00);
\draw[ed] (129.99,-12.00) -- (129.99,-36.00);
\draw[ed] (129.99,-12.00) -- (129.99,-18.00);
\draw[ed] (119.99,-36.00) -- (119.99,-42.00);
\draw[ed] (129.99,-36.00) -- (126.66,-42.00);
\draw[ed] (129.99,-18.00) -- (139.99,-42.00);
\draw[ed] (129.99,-18.00) -- (129.99,-24.00);
\draw[ed] (129.99,-18.00) -- (133.32,-42.00);
\draw[ed] (129.99,-24.00) -- (129.99,-30.00);
\draw[ed] (129.99,-24.00) -- (139.99,-36.00);
\draw[ed] (129.99,-30.00) -- (139.99,-36.00);
\draw[ed] (139.99,-36.00) -- (119.99,-42.00);
\node[ndg] at (119.99,-42.00) {5};
\node[ndg] at (126.66,-42.00) {6};
\node[ndp] at (133.32,-42.00) {12};
\node[ndr] at (139.99,-42.00) {8};
\node[ndg] at (119.99,-36.00) {3};
\node[ndg] at (129.99,-36.00) {4};
\node[ndg] at (139.99,-36.00) {11};
\node[ndg] at (129.99,-30.00) {10};
\node[ndp] at (129.99,-24.00) {9};
\node[ndr] at (129.99,-18.00) {7};
\node[ndr] at (129.99,-12.00) {2};
\node[ndp] at (129.99,-6.00) {1};
\end{scope}
\node[frbadge] at (103.87,-1.20) {frame~20};
\node[status]  at (152.80,-1.20) {6/12 formalized};
\begin{scope}[shift={(\edgecolshift,0)}, shift={(0,\bottomrowgraphshift)}, shift={(24.00,0)}, xscale=\xscale, shift={(-24.00,0)}]
\draw[ed] (24.02,-60.00) -- (24.02,-66.00);
\draw[ed] (24.02,-66.00) -- (12.25,-90.00);
\draw[ed] (24.02,-66.00) -- (18.13,-90.00);
\draw[ed] (24.02,-66.00) -- (24.02,-72.00);
\draw[ed] (12.25,-90.00) -- (12.25,-96.00);
\draw[ed] (18.13,-90.00) -- (15.61,-96.00);
\draw[ed] (24.02,-72.00) -- (35.78,-84.00);
\draw[ed] (24.02,-72.00) -- (24.02,-78.00);
\draw[ed] (24.02,-72.00) -- (18.97,-96.00);
\draw[ed] (35.78,-84.00) -- (35.78,-90.00);
\draw[ed] (35.78,-84.00) -- (22.33,-96.00);
\draw[ed] (35.78,-84.00) -- (25.70,-96.00);
\draw[ed] (35.78,-84.00) -- (32.42,-96.00);
\draw[ed] (35.78,-84.00) -- (35.78,-96.00);
\draw[ed] (24.02,-78.00) -- (12.25,-84.00);
\draw[ed] (24.02,-78.00) -- (24.02,-90.00);
\draw[ed] (12.25,-84.00) -- (24.02,-90.00);
\draw[ed] (24.02,-90.00) -- (12.25,-96.00);
\draw[ed] (35.78,-90.00) -- (25.70,-96.00);
\draw[ed] (29.90,-90.00) -- (29.06,-96.00);
\node[ndg] at (12.25,-96.00) {5};
\node[ndg] at (15.61,-96.00) {6};
\node[ndp] at (18.97,-96.00) {12};
\node[ndp] at (22.33,-96.00) {15};
\node[ndg] at (25.70,-96.00) {16};
\node[ndg] at (29.06,-96.00) {17};
\node[ndp] at (32.42,-96.00) {18};
\node[ndp] at (35.78,-96.00) {19};
\node[ndg] at (12.25,-90.00) {3};
\node[ndg] at (18.13,-90.00) {4};
\node[ndg] at (24.02,-90.00) {11};
\node[ndg] at (29.90,-90.00) {14};
\node[ndr] at (35.78,-90.00) {13};
\node[ndg] at (12.25,-84.00) {10};
\node[ndr] at (35.78,-84.00) {8};
\node[ndp] at (24.02,-78.00) {9};
\node[ndr] at (24.02,-72.00) {7};
\node[ndr] at (24.02,-66.00) {2};
\node[ndp] at (24.02,-60.00) {1};
\end{scope}
\node[frbadge] at (1.20,-46.20) {frame~31};
\node[status]  at (50.13,-46.20) {9/19 formalized};
\begin{scope}[shift={(0,\bottomrowgraphshift)}, shift={(77.00,0)}, xscale=\xscale, shift={(-77.00,0)}]
\draw[ed] (77.00,-60.00) -- (77.00,-64.50);
\draw[ed] (77.00,-64.50) -- (57.00,-91.50);
\draw[ed] (77.00,-64.50) -- (63.67,-91.50);
\draw[ed] (77.00,-64.50) -- (77.00,-69.00);
\draw[ed] (57.00,-91.50) -- (57.00,-96.00);
\draw[ed] (63.67,-91.50) -- (59.86,-96.00);
\draw[ed] (77.00,-69.00) -- (77.00,-73.50);
\draw[ed] (77.00,-69.00) -- (57.00,-82.50);
\draw[ed] (77.00,-69.00) -- (62.71,-96.00);
\draw[ed] (77.00,-73.50) -- (77.00,-78.00);
\draw[ed] (77.00,-73.50) -- (65.57,-96.00);
\draw[ed] (77.00,-73.50) -- (68.43,-96.00);
\draw[ed] (77.00,-73.50) -- (77.00,-96.00);
\draw[ed] (77.00,-73.50) -- (82.71,-96.00);
\draw[ed] (57.00,-82.50) -- (57.00,-87.00);
\draw[ed] (57.00,-82.50) -- (70.33,-91.50);
\draw[ed] (57.00,-87.00) -- (70.33,-91.50);
\draw[ed] (70.33,-91.50) -- (57.00,-96.00);
\draw[ed] (77.00,-78.00) -- (68.43,-96.00);
\draw[ed] (77.00,-78.00) -- (97.00,-82.50);
\draw[ed] (77.00,-78.00) -- (97.00,-87.00);
\draw[ed] (77.00,-78.00) -- (83.67,-91.50);
\draw[ed] (77.00,-91.50) -- (71.29,-96.00);
\draw[ed] (97.00,-82.50) -- (79.86,-96.00);
\draw[ed] (97.00,-82.50) -- (77.00,-87.00);
\draw[ed] (97.00,-82.50) -- (85.57,-96.00);
\draw[ed] (97.00,-82.50) -- (88.43,-96.00);
\draw[ed] (97.00,-82.50) -- (91.29,-96.00);
\draw[ed] (77.00,-87.00) -- (97.00,-91.50);
\draw[ed] (77.00,-87.00) -- (94.14,-96.00);
\draw[ed] (77.00,-87.00) -- (97.00,-96.00);
\draw[ed] (97.00,-87.00) -- (77.00,-91.50);
\draw[ed] (97.00,-87.00) -- (68.43,-96.00);
\draw[ed] (97.00,-87.00) -- (90.33,-91.50);
\draw[ed] (83.67,-91.50) -- (68.43,-96.00);
\draw[ed] (90.33,-91.50) -- (74.14,-96.00);
\draw[ed] (97.00,-91.50) -- (97.00,-96.00);
\node[ndg] at (57.00,-96.00) {5};
\node[ndg] at (59.86,-96.00) {6};
\node[ndp] at (62.71,-96.00) {12};
\node[ndp] at (65.57,-96.00) {15};
\node[ndg] at (68.43,-96.00) {16};
\node[ndg] at (71.29,-96.00) {17};
\node[ndg] at (74.14,-96.00) {20};
\node[ndp] at (77.00,-96.00) {18};
\node[ndg] at (79.86,-96.00) {21};
\node[ndp] at (82.71,-96.00) {19};
\node[ndg] at (85.57,-96.00) {27};
\node[ndp] at (88.43,-96.00) {28};
\node[ndg] at (91.29,-96.00) {29};
\node[ndg] at (94.14,-96.00) {31};
\node[ndg] at (97.00,-96.00) {32};
\node[ndg] at (57.00,-91.50) {3};
\node[ndg] at (63.67,-91.50) {4};
\node[ndg] at (70.33,-91.50) {11};
\node[ndg] at (77.00,-91.50) {14};
\node[ndg] at (83.67,-91.50) {25};
\node[ndg] at (90.33,-91.50) {26};
\node[ndg] at (97.00,-91.50) {30};
\node[ndg] at (57.00,-87.00) {10};
\node[ndg] at (77.00,-87.00) {23};
\node[ndg] at (97.00,-87.00) {24};
\node[ndp] at (57.00,-82.50) {9};
\node[ndp] at (97.00,-82.50) {22};
\node[ndr] at (77.00,-78.00) {13};
\node[ndr] at (77.00,-73.50) {8};
\node[ndr] at (77.00,-69.00) {7};
\node[ndr] at (77.00,-64.50) {2};
\node[ndp] at (77.00,-60.00) {1};
\end{scope}
\node[frbadge] at (52.53,-46.20) {frame~51};
\node[status]  at (101.47,-46.20) {20/32 formalized};
\begin{scope}[shift={(-\edgecolshift,0)}, shift={(0,\bottomrowgraphshift)}, shift={(130.00,0)}, xscale=\xscale, shift={(-130.00,0)}]
\draw[ed] (130.00,-60.00) -- (130.00,-64.50);
\draw[ed] (130.00,-64.50) -- (110.00,-91.50);
\draw[ed] (130.00,-64.50) -- (116.67,-91.50);
\draw[ed] (130.00,-64.50) -- (130.00,-69.00);
\draw[ed] (110.00,-91.50) -- (110.00,-96.00);
\draw[ed] (116.67,-91.50) -- (112.86,-96.00);
\draw[ed] (130.00,-69.00) -- (130.00,-73.50);
\draw[ed] (130.00,-69.00) -- (110.00,-82.50);
\draw[ed] (130.00,-69.00) -- (115.71,-96.00);
\draw[ed] (130.00,-73.50) -- (130.00,-78.00);
\draw[ed] (130.00,-73.50) -- (118.57,-96.00);
\draw[ed] (130.00,-73.50) -- (121.43,-96.00);
\draw[ed] (130.00,-73.50) -- (130.00,-96.00);
\draw[ed] (130.00,-73.50) -- (135.71,-96.00);
\draw[ed] (110.00,-82.50) -- (110.00,-87.00);
\draw[ed] (110.00,-82.50) -- (123.33,-91.50);
\draw[ed] (110.00,-87.00) -- (123.33,-91.50);
\draw[ed] (123.33,-91.50) -- (110.00,-96.00);
\draw[ed] (130.00,-78.00) -- (121.43,-96.00);
\draw[ed] (130.00,-78.00) -- (150.00,-82.50);
\draw[ed] (130.00,-78.00) -- (150.00,-87.00);
\draw[ed] (130.00,-78.00) -- (136.67,-91.50);
\draw[ed] (130.00,-91.50) -- (124.29,-96.00);
\draw[ed] (150.00,-82.50) -- (132.86,-96.00);
\draw[ed] (150.00,-82.50) -- (130.00,-87.00);
\draw[ed] (150.00,-82.50) -- (138.57,-96.00);
\draw[ed] (150.00,-82.50) -- (141.43,-96.00);
\draw[ed] (150.00,-82.50) -- (144.29,-96.00);
\draw[ed] (130.00,-87.00) -- (150.00,-91.50);
\draw[ed] (130.00,-87.00) -- (147.14,-96.00);
\draw[ed] (130.00,-87.00) -- (150.00,-96.00);
\draw[ed] (150.00,-87.00) -- (130.00,-91.50);
\draw[ed] (150.00,-87.00) -- (121.43,-96.00);
\draw[ed] (150.00,-87.00) -- (143.33,-91.50);
\draw[ed] (136.67,-91.50) -- (121.43,-96.00);
\draw[ed] (143.33,-91.50) -- (127.14,-96.00);
\draw[ed] (150.00,-91.50) -- (150.00,-96.00);
\node[ndg] at (110.00,-96.00) {5};
\node[ndg] at (112.86,-96.00) {6};
\node[ndg] at (115.71,-96.00) {12};
\node[ndg] at (118.57,-96.00) {15};
\node[ndg] at (121.43,-96.00) {16};
\node[ndg] at (124.29,-96.00) {17};
\node[ndg] at (127.14,-96.00) {20};
\node[ndg] at (130.00,-96.00) {18};
\node[ndg] at (132.86,-96.00) {21};
\node[ndg] at (135.71,-96.00) {19};
\node[ndg] at (138.57,-96.00) {27};
\node[ndg] at (141.43,-96.00) {28};
\node[ndg] at (144.29,-96.00) {29};
\node[ndg] at (147.14,-96.00) {31};
\node[ndg] at (150.00,-96.00) {32};
\node[ndg] at (110.00,-91.50) {3};
\node[ndg] at (116.67,-91.50) {4};
\node[ndg] at (123.33,-91.50) {11};
\node[ndg] at (130.00,-91.50) {14};
\node[ndg] at (136.67,-91.50) {25};
\node[ndg] at (143.33,-91.50) {26};
\node[ndg] at (150.00,-91.50) {30};
\node[ndg] at (110.00,-87.00) {10};
\node[ndg] at (130.00,-87.00) {23};
\node[ndg] at (150.00,-87.00) {24};
\node[ndg] at (110.00,-82.50) {9};
\node[ndg] at (150.00,-82.50) {22};
\node[ndg] at (130.00,-78.00) {13};
\node[ndg] at (130.00,-73.50) {8};
\node[ndg] at (130.00,-69.00) {7};
\node[ndg] at (130.00,-64.50) {2};
\node[ndg] at (130.00,-60.00) {1};
\end{scope}
\node[frbadge] at (103.87,-46.20) {frame~63};
\node[status]  at (152.80,-46.20) {32/32 formalized};
\end{tikzpicture}

\vspace{1.5mm}
{\footnotesize
\textcolor{nodeformalized}{$\blacksquare$}~formalized \qquad \textcolor{nodegray}{$\blacksquare$}~not yet \qquad \textcolor{nodered}{$\blacksquare$}~currently failing
}
\caption{Evolution of the dependency graph for \textsc{BurnsidePrimeDegreeTheorem} (\textsc{FormalQualBench}). Each panel is a snapshot at one frame; each node represents a statement, and the node map appears in \cref{app:burnside-node-map}. Vertical position shows dependency depth in the displayed graph: base helper nodes appear at level 0, and a node is placed one level above the deepest displayed node it uses. Node color: \textcolor{nodeformalized}{blue} = formalized, \textcolor{nodegray}{gray} = present but not yet, \textcolor{nodered}{red} = currently failing. The node set stabilizes by frame~51, and the final frame closes with all displayed connected statements formalized.}
\label{fig:burnside-evolution}
\end{figure*}
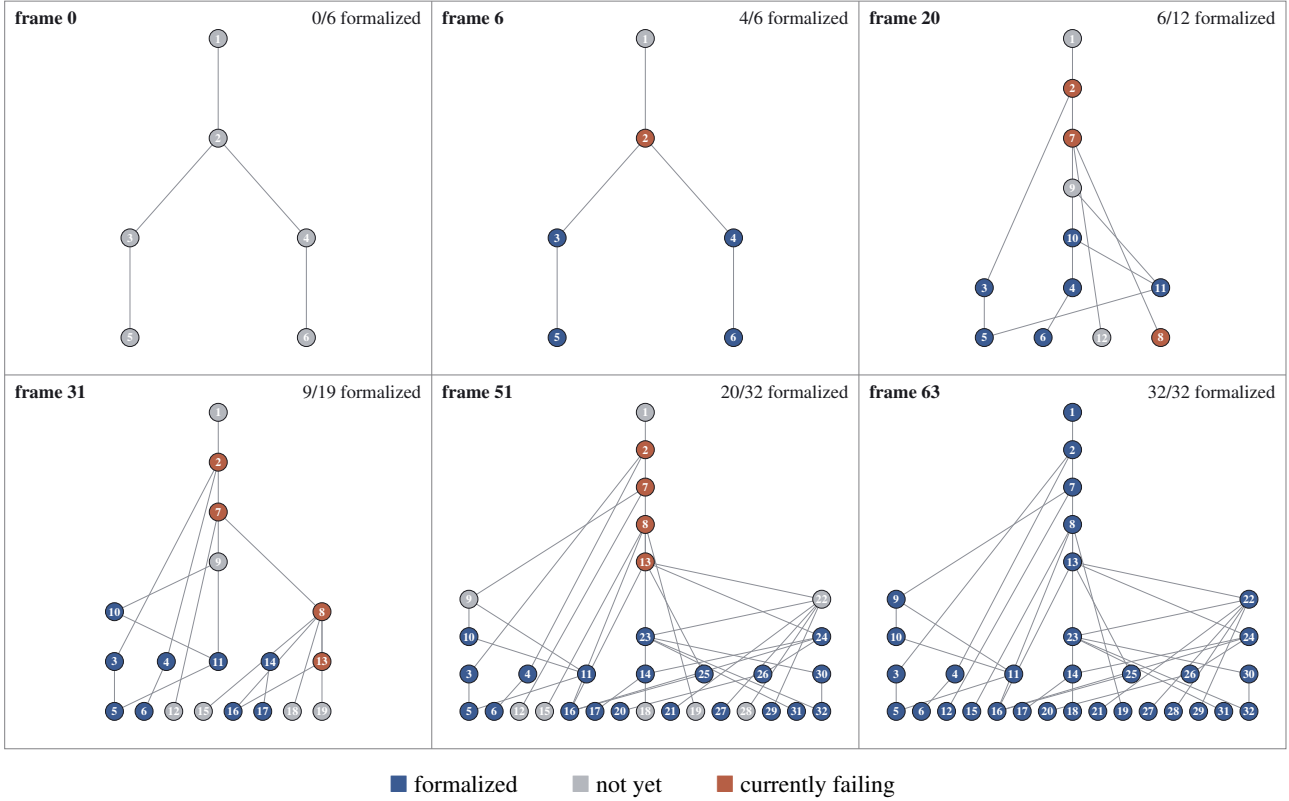